\documentclass{ws-p9-75x6-50}

\begin{document}

\newcommand{\R}{{\bf R}}
\newcommand{\A}{\mathcal{A}}
\newcommand{\Hi}{\mathcal{H}}
\newcommand{\La}{\mathcal{L}}

\title{Supersymmetric Field Theories on Deformed Space-Time }

\author{M. A. Lled\'o}

\address{Dipartimento di Fisica, Politecnico di Torino.\\
Corso Duca degli Abruzzi, 24,\\ 10129 Torino, Italy and\\INFN,
Sezione di Torino, Italy.
\\E-mail: lledo@athena.polito.it}

\maketitle

\abstracts{Field theories on ``quantum" or deformed space-time are
considered here. The Moyal-Weyl deformation breaks the Lorentz
invariance of the theory, but one can still require invariance
under the supertranslation algebra. We investigate some aspects of
the Wess-Zumino model, super Yang-Mills theories and analyze the
correspondence of the later with the supersymmetric Born-Infeld
action. }

\section{Deformation theory: some generalities}

Let $\A$ be the space of $C^\infty$ functions on $\R^m$. A Poisson
bracket on $\A$ is a Lie algebra structure that is a bi-derivation
with respect to the pointwise multiplication, $$\{a,b\cdot
c\}=b\cdot \{a,c\}+\{a,b\}\cdot c, \qquad a,b,c\in \A.$$ We
consider the  standard  symplectic structure ($n=2r$),
$$\{a,b\}=P^{ij}\partial_ia\partial_jb, $$ with $P^{ij}$ a
constant antisymmetric non degenerate matrix.

If $\R^{2n}$ is the phase space of a hamiltonian system, the
quantization map takes (a class of) elements of $\A$ to operators
on the  Hilbert space $\Hi=\La^2(\R^n)$. It was proposed by Dirac
that the quantization map should be such that the Poisson bracket
goes into the Lie bracket of the corresponding operators, that is,
\begin{eqnarray*}D:\A&\longrightarrow &\Theta(\Hi)\\
h\{\;,\;\}&\longrightarrow &[\;,\;].
\end{eqnarray*}
A map like this with reasonable properties for the operators does
not exist (see Ref.\cite{am} for a precise statement of this
fact), although the Poisson bracket of the coordinate functions
may be preserved in the quantization.

The Weyl quantization\cite{w} consists in associating to any
``reasonable" function $a(q,p)$ an operator $A=W(a)$ on
$\La^2({\bf\R}^n)$ defined via the integral $$
A\psi(q)=\frac{1}{(2\pi
h)^n}\int_{\R^{2n}}e^{\frac{i}{h}p(q-q')}a(\frac{q+q'}{2},p)\psi(q')dq'dp,
\quad \psi\in \La(\R^{n}).$$  $a$ is called the Weyl symbol of the
operator $A$.  If $a$ a function in the Schwartz space, the
operator associated by the Weyl map is a bounded operator. If $a$
is a polynomial, the operator is the one obtained by the symmetric
ordering rule.

 Moyal\cite{mo}  wrote the
bracket of such operators in the form,
$$W^{-1}([W(a),W(b)])=h\{a,b\}+\mathcal{O}(h^2).$$ The first term
is the Poisson bracket and the remaining terms are necessary
corrections to Dirac quantization.

The composition formula gives the star product on
$C^\infty(\R^{2n})$. If $A=W(a)$ and $B=W(b)$, then the Weyl
symbol of $A\circ B$ is the associative, non commutative  star
product $$a\star b=a\cdot b+\sum_{n=1}^{\infty}h^kP^k(a,b), \qquad
\qquad{\rm where}$$ $$ P^k(a,b)=P^{i_1j_1}P^{i_2j_2}\cdots
P^{i_kj_k}(\partial_{i_1}\partial_{i_2}\dots
\partial_{i_n})a\cdot(\partial_{j_1}\partial_{j_2}
\dots\partial_{j_k}b).$$ The star product is given as a series in
$h$. There is no guarantee that such series is pointwise
convergent, so the star product is well defined only on the space
of formal power series in $h$ with values in $C^\infty(\R^{2n})$,
namely $C^\infty(\R^{2n})[[h]]$. It was in fact shown in
Ref.\cite{ru} that no associative, non commutative star product
converges for all $C^\infty(\R^{2n})$. One obtains subalgebras
that converge for the Schwartz space and polynomials, as we said
before. For more general classes of symbols, the star product
obeys only  an asymptotic convergence (see for example
Ref.\cite{fe}).

For any star product, the associativity condition at first and
second orders in $h$ assures that $$\lim_{h\mapsto 0}\frac{a\star
b-b \star a}{h}$$ is a Poisson bracket. Once the Poisson bracket
in $\R^{2n}$ is fixed there is, up to isomorphism,  only one  star
product.

\section{Deforming superspace}

The super space of dimension $(p,q)$ is the affine space $\R^p$
together with a
 commutative super algebra

 $$\mathcal{S}^{p,q}=C^\infty(\R^p)\otimes\Lambda(\R^q)=\{a_0(x)+a_i(x)\theta^i+a_{i_1i_2}\theta^{i_1i_2} +\cdots
a_{i_1\dots
 i_q}\theta^{i_1}\wedge\theta^{i_q}\}.$$
 On $S^{p,q}$, left and right, odd and even derivations can be
 defined.  The expression
$$ \{\Phi,\Psi\}=P^{ab}\partial_a\Phi\partial_b\Psi + P^{\alpha
\beta}\partial^R_\alpha\Phi\partial^L_\beta\Psi=
P^{AB}\partial^R_A\Phi\partial^L_B\Psi,  $$ where $P^{\mu\nu}$ and
$P^{ij}$  are constant matrices, antisymmetric and symmetric
respectively, defines a Poisson bracket on $\mathcal{S}^{p,q}$.
The superindices $L$ and $R$ denote left and right derivations
respectively.

A star product on $\mathcal{S}^{p,q}$ is defined as the Weyl
quantization of odd variables\cite{bm}. The expression of the star
product is $$ e^{hP}=\sum_{n=0}^\infty \frac{h^n}{n!}P^n,\qquad
\qquad {\rm with} $$
 \begin{equation} P^n(\Phi\otimes
\Psi)=P^{A_1B_1}P^{A_2B_2}\cdots
P^{A_nB_n}(\partial^R_{A_1}\partial^R_{A_2}\dots
\partial^R_{A_n})\Phi\cdot(\partial^L_{B_1}\partial^L_{B_2}
\dots\partial^L_{B_n}\Psi).\label{spb}\end{equation} If we
consider a deformation only of the odd part of the superalgebra,
$\Lambda(\R^q)\}$, the algebra that one obtains is isomorphic to a
Clifford algebra\cite{bm,fl}. Clifford algebras are then non
commutative superalgebras.

If the manifold that we are deforming is space-time, one can ask
for the behaviour of the star product under super Poincar\'e
transformations. Lorentz transformations are not automorphisms of
the algebra, while translational invariance is preserved. If
$P^{ij}$ is constant, the star product does not behave well under
supertranslations. Using the covariant derivatives
$D^{R,L}_\alpha,\bar D^{R,L}_{\dot\alpha}$, one can define a new
Poisson bracket\cite{fl},
\begin{equation}
\{\Phi,\Psi\}=P^{\mu\nu}\partial_\mu\Phi\partial_\nu\Psi
+P^{\alpha \beta } {D^{R}}_{\alpha }\Phi{D^{L}}_{\beta }\Psi.
\label{spb2}
\end{equation}
The star product can be defined via the exponential as in
(\ref{spb}). We note that the Poisson bracket is always degenerate
in the space of odd variables because of the non trivial
commutation relations among $D$ and $\bar D$. The construction is
easily extended to  $N$ supersymmetries by using harmonic
superspace.

Finally we notice that chiral superfields $\bar D\Phi=0$ are not a
subalgebra of the star product unless $P^{\alpha\beta}=0$.

\section{Supersymmetric deformed field theories}

We will consider only deformations only of the even part of the
superalgebra. If $\Phi_1$ and $\Phi_2$ are two superfields we have
that $$\int d^4x\phi_1\star \Phi_2=\int d^4x\phi_1\cdot\Phi_2=\int
d^4x\phi_2\star \Phi_1,$$ if $\partial_{\mu_1}\cdots
\partial_{\mu_n}\Phi\mapsto 0$ when $x\mapsto \infty$ for all $n$.
Notice that these boundary conditions are not enough to assure the
convergence of the star product.

Let $\Phi$ be a chiral superfield with the expansion $$
\Phi=A(y)+\sqrt{2}\theta\psi(y) +\theta\theta F(y). $$ As a first
xample one can consider the Wess-Zumino model, with action
$$S_{DWZ}= \int d^4xd^2\theta d^2\bar\theta\; \Phi\bar\Phi+\int
d^4x\;(\int d^2\theta \;(\frac{m}{2}\Phi^2 +
\frac{g}{3}\Phi^{\star 3}) +\mbox{c. c.}).$$ The auxiliary field
$F$ satisfies algebraic equations $$ F=-m\bar A-g\bar A\star\bar
A,$$ so the quartic potential becomes $(A\star A)(\bar A\star \bar
A)$ as opposed to the other possible generalization, $(A\star\bar
A)^2$.
\subsection{Rank 1 gauge theory on deformed superspace}
Even for the rank 1 theory the gauge symmetry is non abelian, so
one has to introduce the formalism of non abelian supersymmetric
Yang-Mills theories\cite{fz}. The  elements of the gauge group in
deformed superspace are complex chiral superfields $$ U=e^{\star
i\Lambda}=\sum_{n=0}^\infty \frac{1}{n!}(i\Lambda)^{\star n},
$$with group law $U_1\star U_2=U_3$.

The connection superfield is $e^{\star V}$ and transforms as
$$e^{\star V}\longrightarrow U^\dagger\star  e^{\star V}\star U$$
while the chiral field strength, $W_\alpha={\bar D}^2e^{\star
-V}\star D_\alpha e^{\star V}$ transforms as
$$W_\alpha\longrightarrow U^{\star-1}\star W_\alpha \star U.$$

The deformed super Yang-mills action is $$S_{DSYM}=\int d^4x\int
d^2\theta\; W_\alpha\star W^\alpha \;+\; {\rm c. c.}$$ One can
choose a Wess-Zumino gauge, $V^{\star 3}=0$. It is not preserved
by supersymmetry, and in addition, depends explicitly on the
deformation parameter. The supersymmetry algebra is then realized
in the subset of fields satisfying the Wess-Zumino condition in a
way that depends on the deformation parameter, but the
supersymmetry algebra is  not deformed.

\subsection{Connection with open string theory}
Non commutative gauge fields appear in the context of string
theory as a way to incorporate a vacuum expectation value for the
$B$ field in the effective theory\cite{cds}.

The effective action of open string at low momenta is the
Dirac-Born-Infeld action
$$\La_{DBI}=\sqrt{\det(g_{\mu\nu}+B_{\mu\nu}+F_{\mu\nu})}.$$ It
was argued in Ref.\cite{sw} that one can use an alternative
description in terms of non commutative gauge fields where all the
dependence on $B$ is encoded in the star product. In fact, non
abelian gauge theories in the canonical formalism have an algebra
of first class constrains (which generate the gauge group)
\begin{equation}\{\phi_i,\phi_j\}=c_{ij}^k\phi_k\label{alge}\end{equation} where $\{\;,\;\}$ is the
Poisson bracket. $\{\phi_i\}$ are constrains defining a
submanifold on the phase space. The same submanifold can be
described with a different set of constrains $\{\phi_i\}$ and the
Poisson bracket relations \ref{alge} may not be preserved. In
particular, it was shown by Batalin adn Fradkin\cite{bf} that when
the phase space has a finite number of degrees of freedom, the
algebra can be brought to be abelian, $$\{\phi'_i,\phi'_j\}=0.$$
This was called {\it abelization} of the gauge algebra. In field
theory the abelization can of course introduce a non local change
in the fields.

Seiberg and Witten\cite{sw} found an explicit change of variables
which performs the abelization of the system. If $A$ is the
connection field and $\lambda$ the gauge parameter, the
transformation is of the form $$A\longrightarrow \hat A(A)$$
$$\lambda\longrightarrow \hat\lambda(A,\lambda),$$ where the new
variables are ordinary U(1) gauge fields.  In Ref.\cite{fl} we
showed that this change of variables is consistent with
supersymmetry.

In the limit $\alpha'\mapsto 0$ one can check some properties of
the commutative and the non commutative actions. In Ref.\cite{sw},
the Dirac-Born-Infeld action for commutative and non commutative
fields where compared. The supersymmetric actions can be compared
using the Cecotti-Ferrara\cite{cf} formalism. The action in terms
of the non commutative superfields  is quadratic in this limit,
$$\int d^4x\int d^2 \theta\;\hat W_\alpha\star {\hat W}^\alpha.$$
It has a non linear fermionic symmetry despite the fact that the
action is not free, $$\delta W_\alpha=\eta_\alpha.$$ Since this
theory is supposed to be equivalent to the supersymmetric
Dirac-Born-Infeld theory in this limit, the symmetry of the non
commutative action may correspond to the spontaneously broken
supersymmetry that appears in the commutative one ($N=2$ broken to
$N=1$ supersymmetry).

\section*{Acknowledgments}
I want to thank my collaborator S. Ferrara for his help and
support.

\end{document}